\documentclass[conference]{IEEEtran}
\IEEEoverridecommandlockouts

\usepackage{cite}
\usepackage{amsmath,amssymb,amsfonts}
\usepackage{algorithmic}
\usepackage{graphicx}
\usepackage{textcomp}
\usepackage{xcolor}    
\def\BibTeX{{\rm B\kern-.05em{\sc i\kern-.025em b}\kern-.08em
    T\kern-.1667em\lower.7ex\hbox{E}\kern-.125emX}}
    
    \def\figref#1{Fig.~\ref{#1}}

\makeatletter
\def\ps@IEEEtitlepagestyle{%
  \def\@oddfoot{\mycopyrightnotice}%
  \def\@evenfoot{}%
}
\def\mycopyrightnotice{%
  {\footnotesize  \hfill}% <--- Change here
  \gdef\mycopyrightnotice{}% just in case
}
\makeatother

\begin{document}

\title{Validation of the Heuristic Model for the Electron Mobility in Dense
Helium Gas}
\author{\IEEEauthorblockN{%1\textsuperscript{st} 
Armando Francesco Borghesani}
\IEEEauthorblockA{\textit{CNISM Unit-Department of Physics \& Astronomy} \\
\textit{Università degli Studi di Padova}\\
Padua, Italy \\
armandofrancesco.borghesani@unipd.it}}

\maketitle

%%%%%%%%%%%%%%%%%%%%%%%%%%%%%%%%%%%%%%%%%%%%%%%%%%%%%%%%%%%%%%%%%%%%%%%%%%%

\begin{abstract}
  The investigation of the mobility \(\mu\) of quasi-free electrons in dense, cold
  helium gas as a function of its density \(N\) is made difficult by the fact that a significant fraction of electrons
  is localized in low mobility bubbles. The measured mobility is a weighted
  average of the contributions of the slow electron bubbles, whose mobility is
  determined by hydrodynamics, and of the fast quasi-free electrons. A precise
  description of the fast electron mobility is thus required. In the past we
  have developed a heuristic model that proved succesful at rationalizing the
  electron mobility in argon and neon gases. In order to validate it, we have
  carried out new accurate %electron mobility 
  measurements in helium gas in the
  temperature range $\textbf{26} \, \text{K} < \text{\em T} < \textbf{300}
  \,$K at densities, in which the presence of localized
  states is negligible. The analysis of these new data confirms the validity
  of the model.
\end{abstract}


\section{Introduction}
The electron transport in dense disordered systems is an actively investigated topics since many years. Beside the obvious interest for technical applications, the study  of the electron drift under the action of an externally applied electric field gives the researchers many pieces of information concerning, among others,  the electron-atom scattering properties, the dynamics, and the energetics of electron states in systems which are not amenable to a simple description as that based on the concept of conduction bands in the solid state.  

The behavior of excess electrons has extensively been studied in noble gases because the spherical symmetry of their electronic shells makes quite easy the description of the electron-atom interaction. In spite of the simplicity of such systems,
in a dense noble gas the electron drift cannot simply be described within the binary collision approach of classical kinetic theory because the spatial extension of the electron wave packet leads to a simultaneous interaction with a multitude of surrounding atoms so that also the thermodynamic status of the gas has to be taken into account. Actually, the electron density- normalized mobility \(\mu N\), which should be density independent at constant temperature according to classical kinetic theory~\cite{Huxley1974}, shows a negative density effect (i.e., \(\mu N\) decreases with increasing \(N\), in positive scattering length gases, such as helium~\cite{Levine1967a,Bartels1975b,Schwarz1978} and neon~\cite{Borghesani1988,Borghesani1992}, whereas \(\mu N\) increases with increasing \(N\) in negative scattering length gases, such as argon~\cite{Bartels1973b,Borghesani1992a,Borghesani2001a}.

The negative density effect, first discovered in helium~\cite{Levine1967a}, has particularly been a subject of intense experimental and theoretical investigations because of the possible relationship between multiple scattering effects and the electron self-trapping in bubbles induced by the intrinsic disorder of the gas~\cite{Anderson1958,Adams1988,Omalley1992}.

A number of multiple-scattering theories have been developed over the years in order to describe the mobility \(\mu_{0}\) of thermal electrons, i.e., for a vanishingly weak reduced electric field \(E/N\)~\cite{Omalley1980,Braglia1982,Omalley1992,Borghesani2003}. All of them are based on a complex shift of the electron kinetic energy in the dense medium~\cite{Foldy1945,Lax1951} and on quantum corrections to the electron-atom scattering rate when the electron mean free path \(\ell\) and the electron thermal wavelength \(\lambda_{T}\) are comparable~\cite{Polischuk1983}. 

These theories were quite successful at describing the zero-field density-normalized mobility \(\mu_{0}N\) in helium, mainly because the momentum-transfer scattering cross section \(\sigma_\mathrm{mt}\) is practically energy independent~\cite{OMalley1963} but they grossly failed when applied to neon~\cite{Borghesani1988,Borghesani1992} and to argon~\cite{Borghesani1992a,Borghesani2001,Borghesani2001a}, for which they also have to postulate the action of different physical mechanisms. 

Thus, in order to give a unique description of the electron-atom scattering in a dense gas we have developed a heuristic, adjustable-parameters free model that is based on the most relevant multiple-scattering phenomena~\cite{Borghesani1992a}. Three main multiple scattering effects are included in the model: i) the quantum, density-dependent shift \(E_{k}(N)\) of the electron kinetic energy arising from the exclusion of the electrons from the hard-core volume of the atoms~\cite{Fermi1934}, ii) the correlation among scatterers~\cite{Lekner1968}, and iii) the back-scattering rate enhancement along paths connected by time-reversal symmetry~\cite{Ascarelli1992}. The relevance of these effects depends on the energy dependence of \(\sigma_\mathrm{mt}\) and on the thermodynamic state of the gas. 

The influence of effect i) is relevant if \(\sigma_\mathrm{mt}\) strongly depends on the electron energy. Effect ii) is important only if the gas is relatively close to criticality. Finally, effect iii) is dominant if \(\sigma_\mathrm{mt}\) is big and, possibly, energy independent.

This heuristic model has proved successful in argon as well as in neon. In neon \(\sigma_\mathrm{mt}\) is small and rapidly increases  with increasing energy whereas in argon it is large but strongly decreases with increasing energy. 

\begin{figure}[h!]
  \begin{center}
    \resizebox{1\columnwidth}{!}{\includegraphics{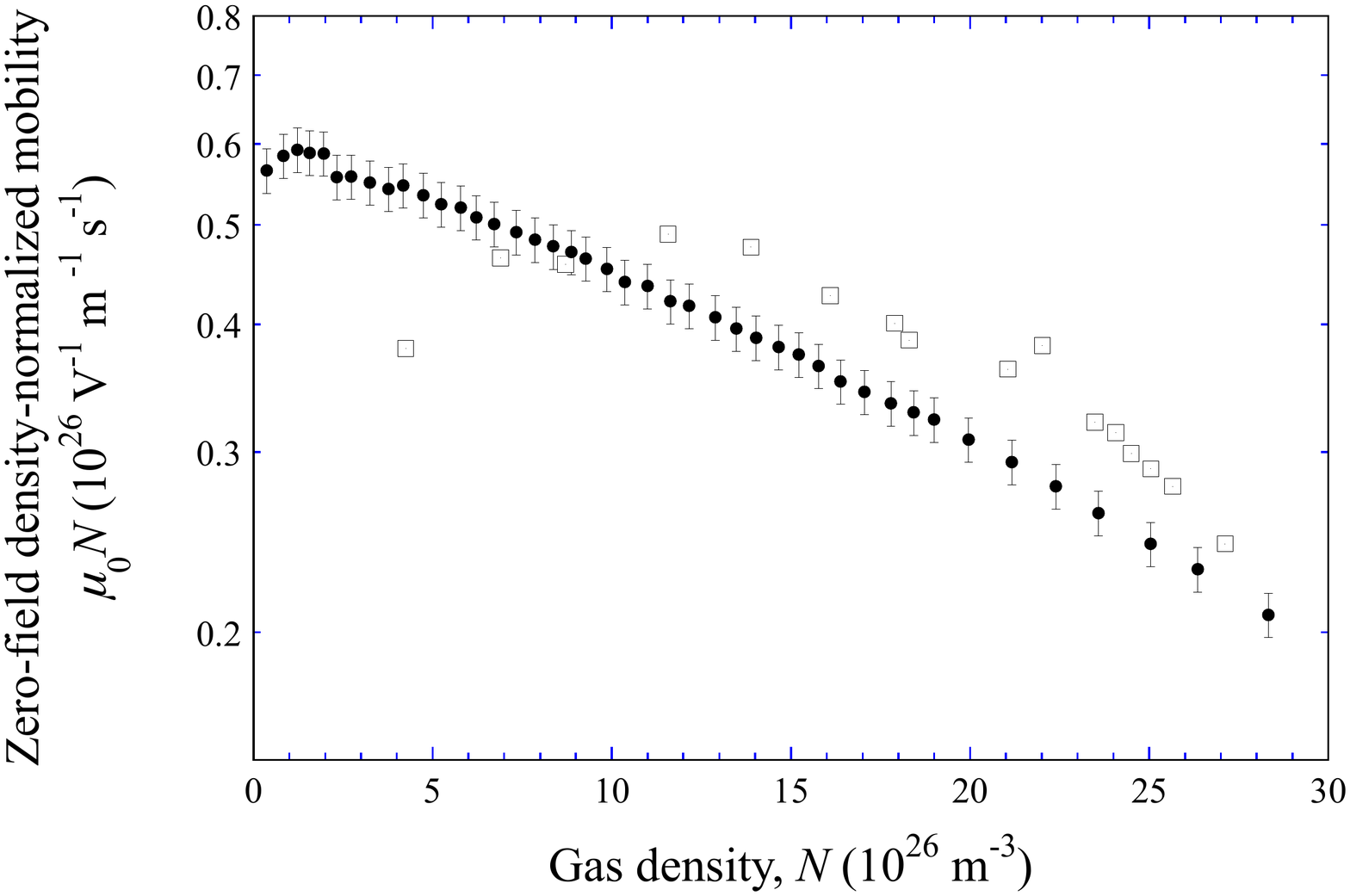}}.
  \end{center}
  \caption{{\small Zero-field density-normalized mobility $\mu_0 N$ as a
  function of the gas density $N$. Open squares: literature data for T=52.8
  K~{\cite{Jahnke1975a}}. Closed points: present experiment data at
  T=54.5 K. \label{fig:confronto}}}
\end{figure}

In order to test its validity we have to prove that it accurately describes the mobility in helium whose \(\sigma_\mathrm{mt}\) is large but almost energy independent. This goal requires that accurate mobility data are available. Unfortunately, in helium, \(\mu_{0}\) was investigated in the past mainly to shed light on the electron localization phenomenon that occurs at pretty high densities whereas the low- and intermediate density regions were not studied with sufficient accuracy, as clearly shown in~\figref{fig:confronto}.  
We have, thus, decided to carry out accurate mobility measurements in a density range in which self-trapping does not occur in a significant way. The different accuracy between the present measurements and typical literature data can be observed in ~\figref{fig:confronto}.
We report here the experimental outcome and the comparison with the heuristic model prediction.

\section{Experimental Results and Discussion}

The experimental technique has thoroughly been described in 
literature~\cite{Borghesani1986,Borghesani1988,Borghesani1990b,Borghesani1990c,Borghesani2001,Borghesani2001a}. We recall here only the main features. The 10 MPa proofed cell is mounted within a thermostat and can be thermoregulated  within \( \Delta T=\pm 0.01\,\)K. The pressure readings have an accuracy \(\Delta P= \pm 10\,\)kPa.
The gas density is computed by means of a literature equation of state~\cite{Sychev1987}. The drift capacitor is powered by a home-made d.c. high-voltage generator~\cite{Borghesani1990b}. The drift distance is \(d=1\,\)cm. A home-made purification system reduces the impurity content to a fraction of a p.p.b.~\cite{Torzo1990}. Bunches of electrons are photoinjected from the cathode by means of a \(4\,\mu\)s short VUV pulse of a Xe flashlamp. The injected charge is in the range \(4\,\mbox{fC}\le Q\le 400\,\mbox{fC}\), depending on the drift field and on the gas density. The current signal is passively integrated in order to improve the signal-to-noise ratio. The drift time \(\tau\) is obtained by the numerical offline analysis of the electron signal and the mobility is obtained as \(\mu=d/\tau E\).

\begin{figure}[h!]
\centering{\includegraphics[width=\columnwidth]{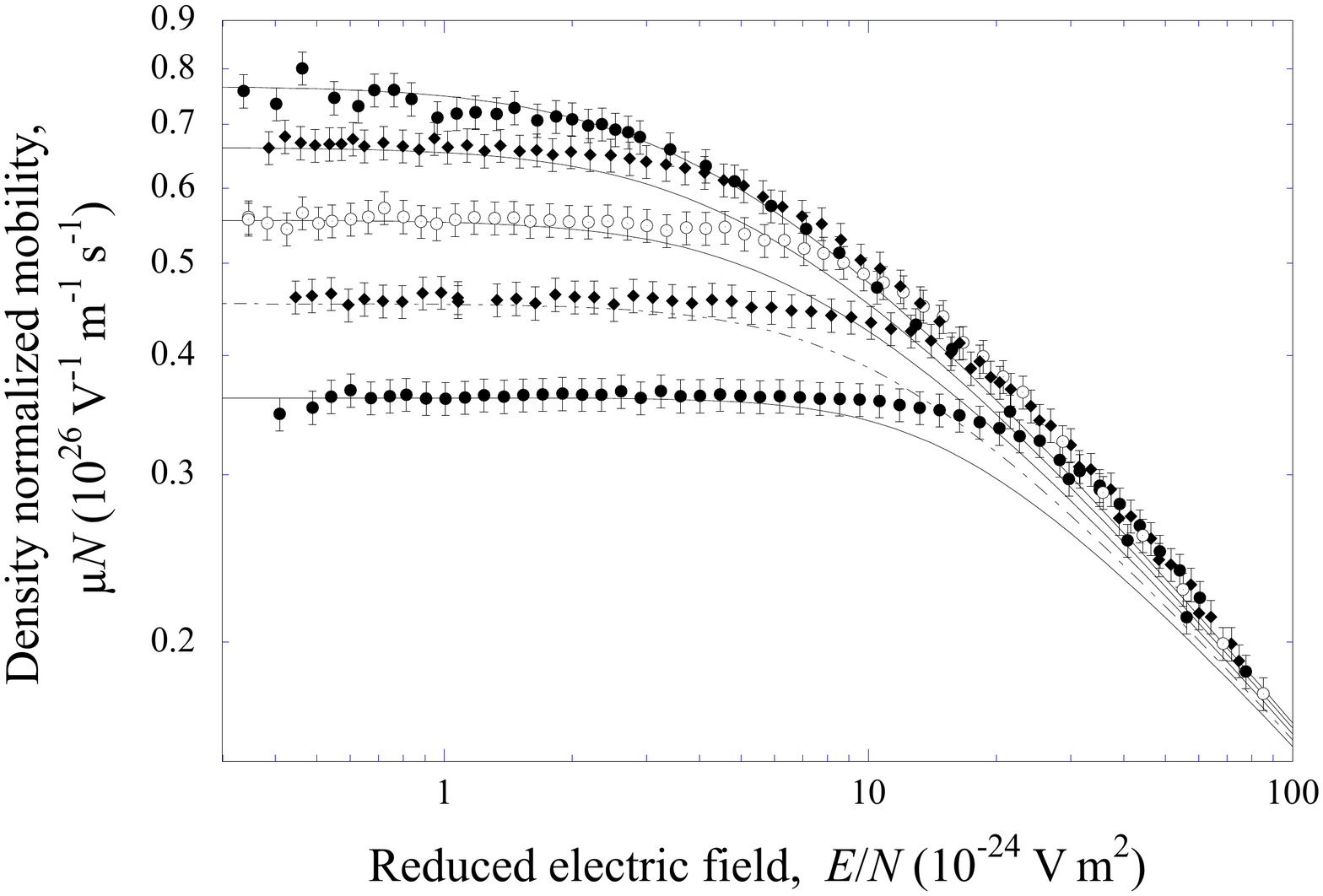}} 
  \caption{{\small Dependence of the density-normalized mobility $\mu N$ on
  the reduced electric field $E / N$ at $T = 34.5 \,$K for
  several densities (in units of $10^{26} \, \text{m}^{- 3}$) $N
  = \,$0.98, 4.62, 8.68, 12.84, 16.98 (from top). The solid lines
  are the predictions of the heuristic model.\label{fig:muNvsEN}}}
\end{figure}

A typical result for \(\mu N\) as a function of \(E/N\) at \(T=34.5\,\)K is shown in \figref{fig:muNvsEN}. The solid lines in the figure are the predictions of the model that will described next. The \(\mu N\) vs \(E/N\) curves are very similar to those  that could be obtained for hard-sphere scattering because \(\sigma_\mathrm{mt}\) is almost independent of the electron energy. At weak field \(\mu N\) is constant whereas it is roughly proportional to \((E/N)^{-1/2}\) at larger field strengths. 

\begin{figure}[h!]
    \centering
    \includegraphics[width=\columnwidth]{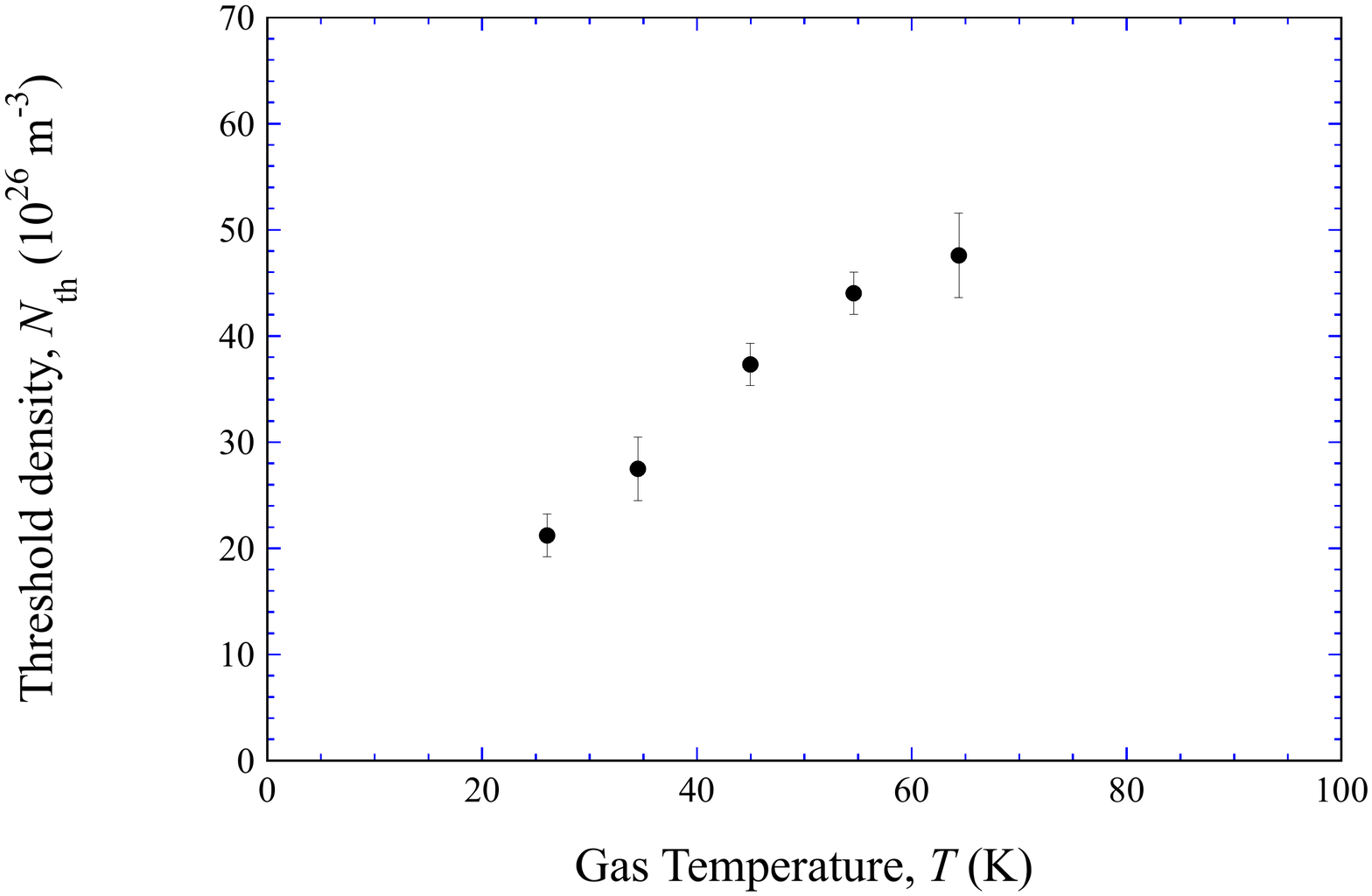}
    \caption{\small Temperature dependence of the threshold density \(N_\mathrm{th}\) below which the electron bubbles concentration is negligible.\label{fig:nth}}
\end{figure}

If electron bubbles were present in a detectable amount \(\mu N\) would show a maximum in between the two limiting behaviors. The threshold density \(N_\mathrm{th} (T)\) above which electron bubbles appear can easily be determined by the analysis of the experimental curves, as done in the neon case~\cite{Borghesani1992}, and is shown in~\figref{fig:nth}. 
We carefully limited the data analysis for \(N<N_\mathrm{th}
\). The data for \(N>N_\mathrm{th}\) are affected by the presence of electron bubbles and have been publisehd elsewhere~\cite{Borghesani2002}. For \(T>65\,\)K the maximum pressure attainable in the experimental cell was not even sufficient to reach \(N_\mathrm{th}.\)

\begin{figure}[h!]
\centering\includegraphics[width=\columnwidth]{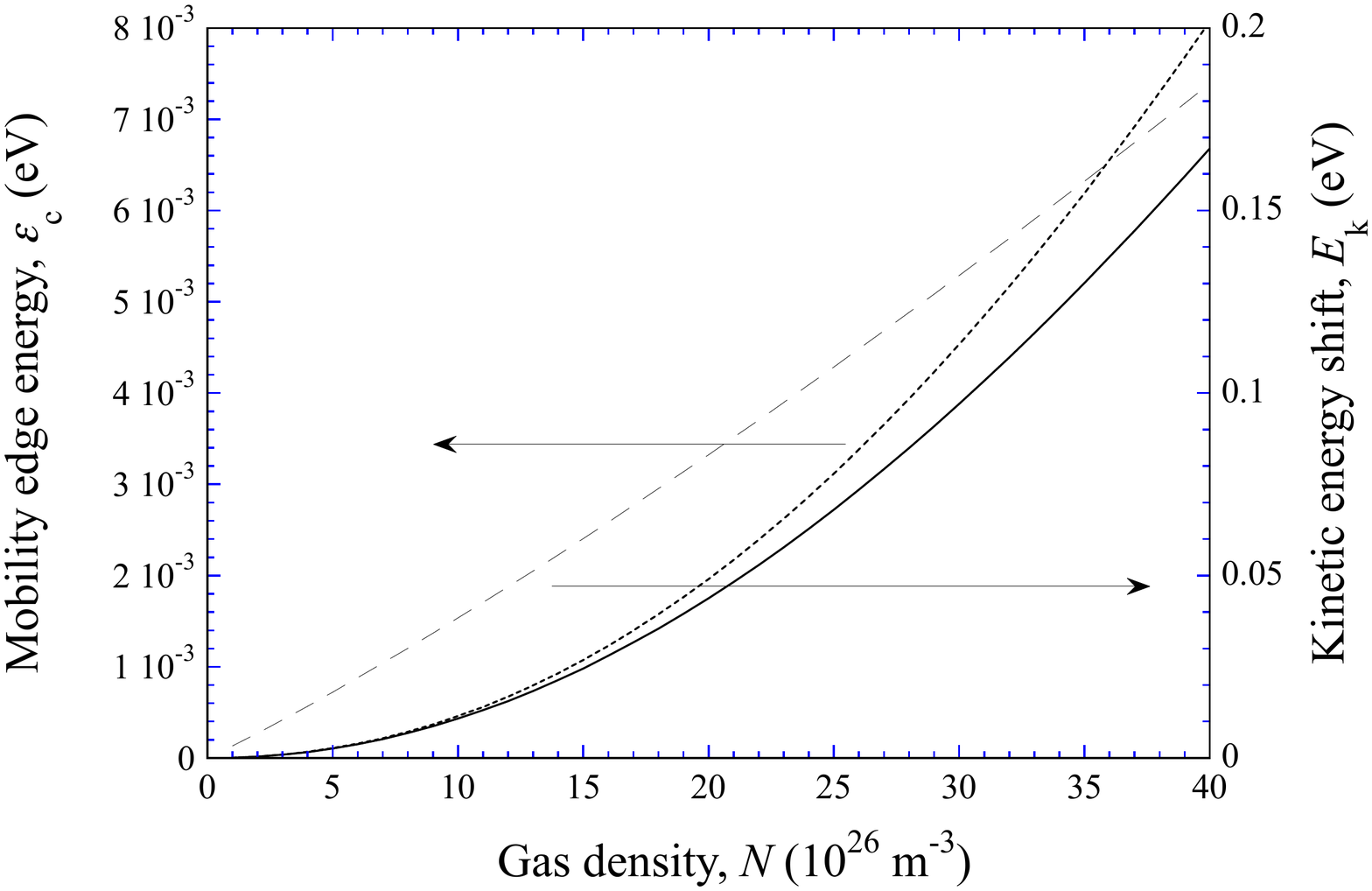} 
  \caption{{\small Density dependence of the kinetic energy shift $E_K$ (right
  scale, dashed line) and of the mobility edge energy $\varepsilon_c$ (left
  scale). $\varepsilon_c$ weakly depends on $T$ because of the static
  structure factor ($T = 26 \,$K: dotted line, $T = 296
  \,$K: solid line).\label{fig:EkEc}}}
\end{figure}

We are now able to set up the equations of the heuristic model. First of all, the excess electron kinetic energy is shifted by the quantum, density-dependent contribution \(E_{k}(N)= (\hbar^{2}/2m)k_{0}^{2}\) in which \(k_{0}\) is the solution of the eigenvalue equation \(
\tan{[k_{0}r_\mathrm{ws}+ \eta_{0}(k_{0})]}=k_{0}r_\mathrm{ws}.
\) \(\eta_{0}(k)\) is the \(s\)-wave phaseshift and \(r_\mathrm{ws}\) is the Wigner-Seitz radius defined by \(4\pi  r_\mathrm{ws}^{3}N=1\). To self-consistently account for the superposition of the atomic potentials \(-\eta_{0}(k_{0})\) is replaced by the hard-core radius of the Hartree-Fock potential \(\tilde a =\sqrt{\sigma_\mathrm{t}(k_{0})/4\pi}\), where  \(\sigma_\mathrm{t}(k)\) is the total scattering cross section. \(E_{k}\) is large in helium because of its large cross section, as shown in~\figref{fig:EkEc} (dashed line, right scale), but has little effect on the mobility because \(\sigma_\mathrm{mt}\) weakly depends on energy.

Correlations among scatterers are accounted for by the static structure factor of the system \(S(k)\) whose long-wavelength limit \(S(0)\) is related to the isothermal gas compressibility \(\chi_{T}\) by \(S(0)=Nk_\mathrm{B}T\chi_{T}\), thereby yielding a proportional increase of the scattering cross section~\cite{Lekner1968}.
In our case these correlations do not significantly influence the mobility because the experiment is carried out at \(T\gg T_{c}\), where \(T_{c}=5.2\,\)K is the helium critical temperature.

The third multiple-scattering effect to be considered is the backscattering rate enhancement due to the quantum self-interference of the electron wave function scattered off atoms located along paths connected by time-reversal symmetry~\cite{Ascarelli1992}. The strength of this effect depends on the ratio \(\lambda/\ell\), where \(\lambda=\hbar/\sqrt{2m\epsilon}\) is the electron quantum wavelength and \(\ell=1/N\sigma_\mathrm{mt}\) is its mean free path. In helium, for some densities and temperatures of the experiment \(\lambda \) may become equal or even greater than \(\ell\), thereby leading to a regime of weak localization~\cite{Adams1988} in which electrons become completely Anderson localized with exponentially decaying wavefunctions and do not contribute to transport. As a consequence, a mobility edge appears at the finite energy \(\epsilon_{c}\) that can suitably be expressed as~\cite{Atrazhev1977,Polischuk1984} 
\begin{equation}
\epsilon_{c}(N,T) =\frac{2}{m}\left\{
\hbar N S(0)\sigma_\mathrm{mt}\left[\epsilon_{c}(N,T)\right]
\right\}^{2}
\label{eq:ec}\end{equation} and is shown in~\figref{fig:EkEc} (dotted and solid lines, left scale). The lower is \(T\), the stronger is the dependence of \(\epsilon_{c}\) on \(N\) because of the influence of \(S(0)\).

The equations of the heuristic model can now be explicitly written down~\cite{Borghesani1992a}. The density-normalized mobility is given by
\begin{equation}
\mu N= -\frac{e}{3}\left(\frac{2}{m}\right)^{1/2}
\int\limits_{\epsilon_{c}}^{\infty}
\left[\frac{\epsilon}{\sigma^{\star}_\mathrm{mt}(\epsilon)}
\right]\frac{\mathrm{d} g(\epsilon)}{\mathrm{d}\epsilon}\,\mathrm{d}\epsilon
\label{eq:mun}\end{equation}
\(\sigma^{\star}_\mathrm{mt}\) is the effective scattering cross secion and
\(g(\epsilon) \) is the Davydov-Pidduck distribution function~\cite{Cohen1967}
\begin{equation}
g(\epsilon) = A \exp{ -\int\limits_{0}^{\epsilon}
\left[
k_\mathrm{B}T + \frac{Me^{2}}{6mz}\left(
\frac{E}{N\sigma^{\star}_\mathrm{mt}(z)}
\right)^{2}
\right]^{-1}
}\,\mathrm{d}z
\label{eq:g}\end{equation} that has to be normalized as \(\int_{0}^{\infty}\epsilon^{1/2}g(\epsilon)\,\mathrm{d}\epsilon =1.\)
The effective scattering cross section is given by
\begin{equation}
\sigma^{\star}_\mathrm{mt}(\epsilon) = \mathcal{F}(w)\sigma_\mathrm{mt}(w)
\left[1-
f\hbar
\frac{\mathcal{F}(w)\sigma_\mathrm{mt}(w)}{(2mw)^{1/2}}
\right]^{-1}
\label{eq:smt}\end{equation} in which \(w=\epsilon+E_{k}(N)\) is the shifted energy and \(\mathcal{F}(w)=(4w^{2})^{-1}\int_{0}^{2w}q^{3}S(q)\,\mathrm{d}q\). We used the  Ornstein-Zernike approximation \(S(q)=[S(0)+(qL)^{2}]/[1+(qL)^{2}]\) with \(L^{2}=10^{-3}[S(0)-1]\,\)nm\(^{2}\)~\cite{Thomas1963}.

\begin{figure}[h!]
   \centering \includegraphics[width=\columnwidth]{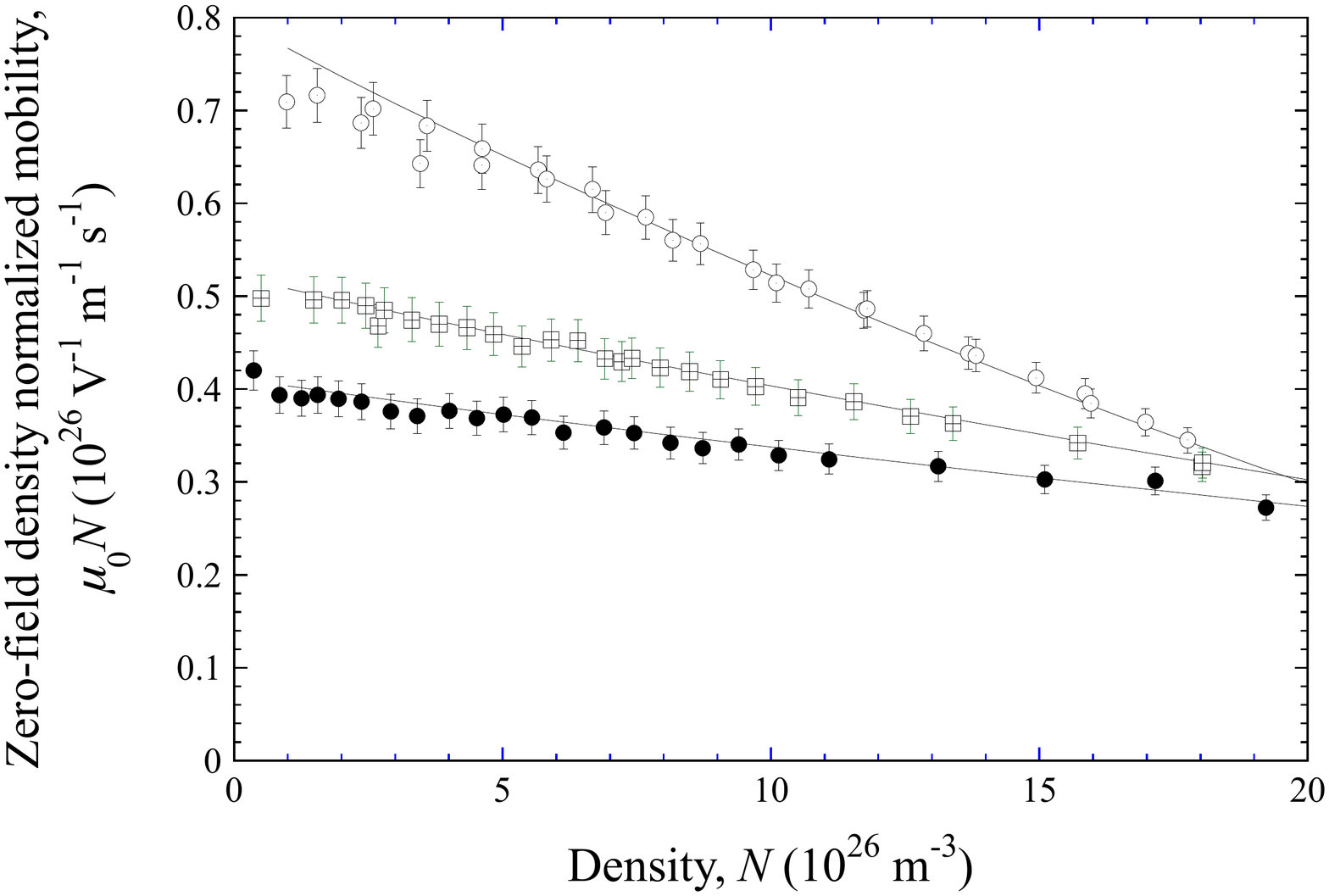}
   \caption{\small Density dependence of the zero-field, density normalized mobility \(\mu_0 N\) for \(T=34.5,\, 77.2\,\mbox{and}\, 120.1\,\)K (from top).   \label{fig:mu0N}}
\end{figure}

In~\figref{fig:mu0N} we now compare the experimental data for the zero-field density-normalized mobility \(\mu_{0}N\) as a function of \(N\) for some \(T\) with the outcome of the heuristic model.
The prediction of the heuristic model (solid lines in the figure) perfectly agrees with the experimental data at all investigated temperatures. Similar results have also been obtained for more temperatures~\cite{Borghesani2021}. It is worth recalling that there are no adjustable parameters in the theory. The specifity of the gas is only accounted for by the scattering cross section and the equation of state. 

We have to note that the influence of the mobility edge \(\epsilon_{c}\) is much stronger than that of the energy shift \(E_{k},\) indeed. \(\mu N\) is sort of a suitable thermal average of \(1/\sigma_\mathrm{mt}(\epsilon)\). Loosely speaking, \(\langle 1/\sigma_\mathrm{mt}(\epsilon)\rangle \approx 1/\sigma_\mathrm{mt}(\langle \epsilon\rangle)\). The action of \(E_{k}\) is to shift \(\langle \epsilon\rangle =(3/2)k_\mathrm{B}T\) to \(\langle \epsilon\rangle =(3/2)k_\mathrm{B}T+ E_{k}\). As the cross section does not depend very much on the energy, the energy shift does not much change the average value of the cross section. On the other hand, \(\epsilon_{c}\) is an infrared cutoff in the integration of a nearly exponentially decaying energy distribution function that cancels the contribution of a significant fraction of highly mobile electron states, thereby leading to the observed decrease of \(\mu_{0}N\) with increasing \(N.\)

The equations of the model also allows us to easily compute the dependence of \(\mu N\) on \(E/N.\) The predictions of the model are shown as solid lines in~\figref{fig:muNvsEN} for \(T=34.5\,\)K. Once more, the agreement with experiment is excellent. Similar results are also obtained for all temperatures not reported here~\cite{Borghesani2021}, provided that \(N_\mathrm{th}\) is not exceeded so that only quasi-free electrons are present.

We can, however, observe a small discrepancy between model and data in the crossover  region between the thermal- and epithermal electron behaviors. We believe that there might be two possible causes for this discrepancy. First of all, the scattering cross section is derived by the analysis of swarm experiments so that there could be an imperfect knowledge of its behavior at energies higher than thermal. Secondly, \(E_{k}(N)\) is obtained as the lowest energy eigenvalue (\(s\)-wave) of the electron kinetic energy operator provided that in the gas there is a local translational symmetry over a distance \(2r_\mathrm{ws}\).%~\cite{Hernandez1991a}. 
 We thus have neglected the possibility that, at stronger \(E/N\), \(E_{k}\) might have some contributions from higher lying eigenvalues of larger angular momentum.

\section{Conclusions}
We have presented new accurate measurements of the mobility of quasi-free electrons in helium gas in density and temperature ranges in which we can rule out the presence of localized electron states that could spoil the interpretation of the experimental data. We are able to rationalize the data with a heuristic model which incorporates some multiple scattering effects  within the frame of the classical kinetic theory  in a heuristic way. The agreement of the model predictions with the experimental data is excellent in spite of the fact that there no adjustable parameters.
We have thus proved that the heuristic model we developd in the past to describe the mobility of quasi-free electrons in argon and neon gas is very well suited also for helium. We thus believe that a unified description of the electron scattering process at low energy in a gas has been achieved.

\bibliographystyle{IEEEtran}

\end{document}